# Critical phenomena of the layered ferrimagnet $Mn_3Si_2Te_6$ following proton irradiation

Rubyann Olmos[1], Jose A. Delgado[1], Hector Iturriaga[1], Luis M. Martinez[1], Christian L. Saiz[1], L. Shao[2], Y. Liu[3], C. Petrovic[4], Srinivasa R. Singamaneni[1*]

[1]Department of Physics, The University of Texas at El Paso, El Paso, Texas 79968, USA

[2]Department of Nuclear Engineering, Texas A&M University, College Station, Texas 77845, USA

[3]Los Alamos National Laboratory, MS K764, Los Alamos NM 87545, USA

[4]Condensed Matter Physics and Materials Science Department, Brookhaven National Laboratory, Upton, New York 11973, USA

## ABSTRACT

In this study, we examine the magnetic properties of the quasi 2D ferrimagnetic single crystal $Mn_3Si_2Te_6$ (MST) through critical phenomena and magnetic entropy analysis in the easy axis (H || *ab*) as a function of proton irradiance. A modified asymptotic analysis method is employed to find the critical exponents $\beta$ and $\gamma$. We find that upon proton irradiation the critical exponents do not fall into any particular universality class but lie close to mean field critical exponents ($\gamma = 1$, $\beta = 0.5$). The presence of long-range interactions can be safely assumed for the pristine and irradiated cases of MST examined in this work. Further analysis on the effective spatial dimensionality reveal that MST remains at $d = 3$ under proton irradiation transitioning from an $n = 1$ spin dimensionality to $n = 2$ and $n = 3$ for $1 \times 10^{15}$ and $5 \times 10^{15}$ $H^+/cm^2$, indicating an *XY* interaction and a Heisenberg interaction, respectively. The pair (spin-spin) correlation function reveals an increase in magnetic correlations at the proton irradiance of $5 \times 10^{15}$ $H^+/cm^2$. In conjunction, the maximum change in magnetic entropy obtained from isothermal magnetization at 3 T is the largest for $5 \times 10^{15}$ $H^+/cm^2$ with a value of 2.45 J/kg K at $T = 73.66$ K, in comparison to 1.60 J/kg K for pristine MST at $T = 73$ K. These results intriguingly align with the findings in previous work where magnetization (at 50 K, 3 T) increased by ~50 % at $5 \times 10^{15}$ $H^+/cm^2$, in comparison to pristine MST [L. M. Martinez et al., Appl. Phys. Lett. **116**, 172404 (2020)]. Magnetic entropy derived from zero-field heat capacity does not show large deviations across the proton irradiated samples. This suggests that the antiferromagnetic coupling between the Mn sites is stable even after proton irradiation. Such result implies that magnetization is enhanced through a strengthening of the super-exchange interaction between Mn atoms mediated through Te rather than a weakening of the AFM component. Overall, our study finds that the magnetic interactions are manipulated the greatest when MST is irradiated at $5 \times 10^{15}$ $H^+/cm^2$.

* srao@utep.edu

## I. INTRODUCTION

Following the discovery of graphene, significant attention has been invested into the study of van der Waals (vdW) materials due to their ability to retain their magnetic properties in low dimensions [1–4]. Moreover, weak interlayer bonding in vdW magnets has made it feasible to chemically synthesize or mechanically exfoliate these materials to one or few layers. This is despite that the Mermin-Wagner theorem prohibits long-range magnetic ordering in an isotropic Heisenberg system at the two-dimensional (2D) limit [5]. Foundationally, the key ingredient to stabilizing the magnetic ordering in vdW magnets is the uniaxial magnetic anisotropy present in these materials [2,6,7]. The emergence of such 2D materials has indeed been realized through the opening of magnon excitation gaps resisting thermal agitations, therefore, allowing constraints imposed by the Mermin-Wagner theory to be lifted [3]. Many 2D materials have been found to be promising candidates for magnetoelectronic, spintronic, data storage, and memory device applications [8–11]. Additionally, bulk magnetoelectrical properties and structural characteristics on the chromium trihalides and transition metal dichalcogenides have been amply studied in the literature since $CrI_3$ and $Cr_2Ge_2Te_6$ were first discovered [2,9]. The ferrimagnetic $Mn_3Si_2Te_6$ (MST) compound, often referred to as a sister magnet to $CrSiTe_3$, has been magnetically characterized in a few studies and has a Curie temperature ($T_C$) between $74 - 78$ K [12–14]. Although there is no orbital moment in MST, there exists a large anisotropy field of 10 T at 5 K [15], despite a $Mn^{2+}$ ion with a spin of 5/2. There are two Mn sites in this compound (Mn1, Mn2) that are not exactly coplanar with a multiplicity of Mn1 being twice that of Mn2. MST has a trigonal crystal structure with space group no. 163 and is composed of $ab$ plane edge sharing $MnTe_6$ octahedra at the Mn1 site (see Figure 1). Layers with vdW gaps are created with Si-Si dimers where the layers are connected by filling up one-third of the octahedral holes with Mn at the Mn2 site [16].

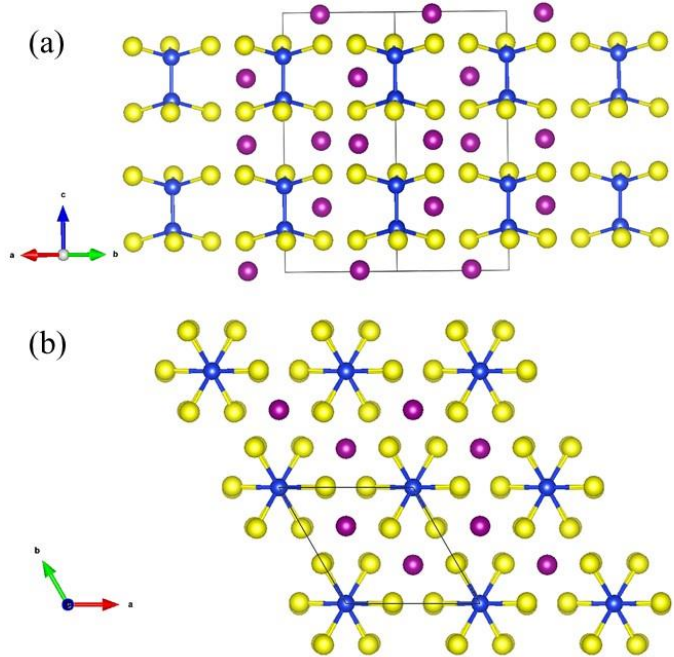

FIG. 1. The crystal structure for MST is seen for (a) the lateral view along the $c$-axis and (b) the top view of the $ab$-plane. The Mn atoms are designated by the purple atom, blue indicates the Si atom, and yellow indicates Te atoms. In (a) the Mn1 and Mn2 sites can be discerned and in (b) the octahedra forming MnTe6 geometry can be seen where the black rectangle indicates the unit cell in both figures.

Today, 2D materials are being rigorously modified to control their magnetoelectric states by employing external perturbations through chemical, mechanical, and electrical incitements. For example, electrostatic doping has shown to modify magnetic properties through the manipulation of the exchange parameters and magnetic anisotropy, and chemical doping has also tuned the magnetic and optical properties of these class of materials [17,18]. Proton irradiation has been shown to significantly alter the magnetic and electronic properties of various materials including $MoS_2$, graphene, and other layered materials [19–26]. In our previous work [14], magnetization was significantly enhanced by 53% and 37% in the ferrimagnetic phase (at 50 K) when MST was irradiated with protons at the fluence of $5 \times 10^{15}$ H$^+$/cm$^2$, along the $ab$ and $c$ plane, respectively. Moreover, analytically studying the critical phenomena can be used to discern the behavior of a materials physical system at its critical point. The importance of critical behavior analysis has recently been demonstrated on the vdW material $Cr_2X_2Te_6$ (X = Si, Ge), where 2D-Ising systems were identified and coupled with long-range interactions [27,28]. Upon studying the anisotropic critical behavior of these materials, it was revealed that the potential proponent for sustaining long-range interactions was magnetocrystalline anisotropy [12].

In this study, the critical phenomena of proton irradiated MST are analyzed to investigate how the magnetic properties of MST evolve as a function of proton irradiance for measurements in the easy axis (H ∥ $ab$). Measurements were conducted on five samples: a non-irradiated pristine crystal and four crystals irradiated at the following proton fluence rates: $1 \times 10^{15}$, $5 \times 10^{15}$, $1 \times 10^{16}$, and $1 \times 10^{18}$ H$^+$/cm$^2$. A modified asymptotic analysis method developed by S. N. Kaul [29] was employed to conduct the critical exponent analysis and is often used as a tool to characterize the magnetic properties of novel materials. Critical exponents for pristine and proton irradiated MST were found to exhibit an increasingly mean field-like behavior as a function of proton fluence. The reliability of these critical exponents is confirmed by performing a scaling analysis. Analysis on the pair correlation function revealed a possible spin transition where spin dimensionality goes from $n = 3, 2,$ and 1 for pristine, $1 \times 10^{15}$, and $5 \times 10^{15}$ H$^+$/cm$^2$, respectively. The magnetic entropy and heat capacity are also examined to elucidate how the magnetic properties were influenced after utilizing protons to perturb the pristine magnetic state. The change in magnetic entropy revealed a maximum for the fluence of $5 \times 10^{15}$ H$^+$/cm$^2$, corresponding to the largest increase in magnetization as reported earlier [14]. Heat capacity measurements confirmed the $T_C$ for all proton irradiated samples, where little variation of $T_C$ exists between the array of samples studied. Moreover, an analysis from heat capacity is carried out by calculating the magnetic entropy by performing a linear background subtraction and Einstein fits in two separate analyses.

## II. METHODOLOGY

### A. Experimental

Bulk MST crystals were synthesized as outlined in a previous report by co-authors, Y. Liu and C. Petrovic [13]. Magnetization and heat capacity measurements were performed using a Quantum Design PPMS VersaLab system with a temperature range of 50 – 400 K and magnetic field of ± 3 T. Magnetization measurements were performed with a vibrating sample magnetometer (VSM) with the magnetic field applied along the $ab$ plane for each the pristine and irradiated sample. MST samples were irradiated with protons at Texas A&M University. A 2 MeV proton beam and a 1.7

MV Tandetron accelerator was utilized at $1 \times 10^{15}$, $5 \times 10^{15}$, $1 \times 10^{16}$, and $1 \times 10^{18}$ H+/cm². Further details on the proton irradiation process are described in previous reports [30,31].

B. Critical Exponent Analysis

To evaluate the critical parameters of MST as a function of proton fluence, a modified asymptotic analysis approach is employed. In a review by S. N. Kaul [29], the modified asymptotic analysis method employed in this work was developed with the goal of arriving at the most accurate critical parameters. In the modified asymptotic analysis method, an appropriate spin model is used to construct a modified Arrott plot (MAP) ($M^{1/\beta}$ vs. $(H/M)^{1/\gamma}$). From the MAP, spontaneous magnetization, $M_S(T)$, and inverse initial susceptibility, $\chi_0^{-1}(T)$, are obtained from the intercepts of the $M^{1/\beta}$ and $(H/M)^{1/\gamma}$ axes, respectively. Fitting of equations (1) and (2) reveal the $T_C$ associated with $\beta$ and $\gamma$. With the intercepts ($M_S(T)$, $\chi_0^{-1}(T)$), the Kouvel-Fisher (KF) plot is generated and the critical exponents $\beta$ and $\gamma$ are extracted and used to construct a new MAP. This process is then repeated iteratively until the exponents converge or remain unchanged. An automated program described in Ref. [32] is used to perform self-consistent calculations as outlined above. Throughout the iterative process the MAP isotherms in the critical regime should become increasingly parallel and straight. Although there are inherent difficulties extrapolating the values of critical exponents and $T_C$, the values extracted using the modified asymptotic analysis are more reliable and converge more quickly compared to other asymptotic analysis methods [29].

A second order phase transition from a paramagnetic (PM) to ferromagnetic (FM) state can be characterized by the critical exponents $\beta, \gamma, \delta, \eta, \nu$. The exponents are determined in terms of the dimensionless variable, $\varepsilon = (T - T_C)/T_C$, known as the reduced temperature, and the critical amplitudes $M_0$ and $(h_0/m_0)$ [33]. The magnetization exponents $\beta, \gamma$, and $\delta$, are characterized by the scaling hypothesis and are obtained using the following power law equations:

$$M_S(T) = M_0(-\varepsilon)^\beta, \quad \varepsilon < 0, T < T_C, \tag{1}$$

$$\chi_0^{-1}(T) = \left(\frac{h_0}{m_0}\right)\varepsilon^\gamma, \quad \varepsilon > 0, T > T_C, \tag{2}$$

$$M = DH^{1/\delta}, \quad T = T_C, \tag{3}$$

To determine the critical exponents associated with the previous equations, the KF plot and Arrott plot techniques are employed to extract $\beta$ and $\gamma$ from $M_S(T)$ and $\chi_0^{-1}(T)$, respectively. The critical isotherm (Eq. (3)) can be fit to extract an effective value for $\delta$. The scaling hypothesis also gives the magnetic equation of state in the critical region, and can be expressed as:

$$M(H, \varepsilon) = \varepsilon^\beta f_\pm\left(\frac{H}{\varepsilon^{\beta+\gamma}}\right), \tag{4}$$

where the regular functions $f_+$ and $f_-$ correspond to $T > T_C$ and $T < T_C$, respectively. Defining renormalized magnetization, $m$, and field, $h$, as $m \equiv \varepsilon^{-\beta}M(H, \varepsilon)$ and $h \equiv \varepsilon^{-(\beta+\gamma)}H$, the equation of state transforms to $m = f_\pm(h)$. This implies that for true scaling relations and proper selection of $\beta, \gamma$, and $\delta$, that scaled $m$ and $h$ will fall onto two universal curves, one below and one above $T_C$.

## III. RESULTS AND DISCUSSIONS

### A. Magnetization and critical phenomena

In a previous work, the magnetization of MST decreased as a function of proton irradiation except at 5 x $10^{15}$ H$^+$/cm$^2$ where the magnetization was significantly enhanced by 53% and 37% for magnetic fields parallel to the $ab$ and $c$ planes, respectively [14]. Figure 2(a) shows isothermal magnetization for this fluence with temperature isotherms in the vicinity of the $T_C$ between 60 to 80 K. The easy axis (H ∥ $ab$) is initially confirmed through isothermal magnetization for the control measurements carried out on the pristine MST sample [14]. Additional first quadrant isothermal magnetization data for other proton fluences has been provided in the Supplemental Material in Figure S1 [34].

The general behavior of MST critical exponents is verified through a preliminary analysis shown in Supplemental Material Figure S2 [34]. An Arrott plot ($M^2$ vs. $H/M$) is constructed using mean field critical exponents ($\beta = 0.5$ and $\gamma = 1.0$) producing a set of straight parallel lines with the isotherm at $T_C$ passing through the origin [35]. Pristine isothermal curves in the Arrott plot exhibit a nonlinear behavior with downward curvature from low to high fields. Therefore, the mean field model does not entirely apply to MST for the pristine crystal where parallel lines were obtained for $\beta = 0.41$ and $\gamma = 1.21$, as implicated in Ref. [13]. Furthermore, the order of the magnetic transition can be estimated through the criterion established by Banerjee by examining the slope of the straight lines in the Arrott plot [36]. A negative slope corresponds to a first-order transition and a positive slope to a second order transition, with the latter recently confirmed for pristine in Ref. [13] and in this work, similarly for the irradiated set MST samples.

To further characterize proton irradiated MST, we adopt the MAP technique which is constructed using the Arrott-Noakes equation of state:

$$\left(\frac{H}{M}\right)^{1/\gamma} = a\varepsilon + bM^{1/\beta},\qquad(5)$$

where $a$ and $b$ are constants and $\varepsilon$ is the reduced temperature, which has been previously defined in Ref. [38]. The MAP as seen in Figure 2(b) for $5 \times 10^{15}$ H$^+$/cm$^2$ MST shows a clear set of parallel isotherms in the high field region. Figure 2(c) shows the spontaneous magnetization and initial inverse susceptibility plots from Eqs. (1) and (2) for the fluence $5 \times 10^{15}$ H$^+$/cm$^2$. From these final power law plots the $T_C$ associated with the critical exponent $\beta$ and $\gamma$ are extracted and listed in Table I for all samples.

Table I. Critical exponents for pristine and proton irradiated Mn$_3$Si$_2$Te$_6$ for different techniques such as Kouvel-Fisher (KF), neutron diffraction, and theory. $T_C$ as seen in fourth column of this table is extracted from the power law dependence of $M_S$ and $\chi_0^{-1}$ exponents, relating to beta and gamma, respectively. This is following the asymptotic analysis method as spelled out in S. N. Kaul [27]. The experimental critical isotherm exponent delta is calculated through the Widom scaling law relation [35]. A visual of how beta,

gamma, delta, and $T_C$ behave as a function of fluence can be seen in Figure S6 of the supplemental information.

| Composition | Reference | Technique | $T_C$ (K) | $\beta$ | $\gamma$ | $\delta$ |
|---|---|---|---|---|---|---|
| Pristine $Mn_3Si_2Te_6$ | This work | KF | 73.46, 73.36 | 0.41(9) | 1.11(0) | 3.65(2) |
| $1 \times 10^{15}$ ($H^+$/cm$^2$) $Mn_3Si_2Te_6$ | This work | KF | 73.20, 73.22 | 0.45(2) | 1.06(9) | 3.36(3) |
| $5 \times 10^{15}$ ($H^+$/cm$^2$) $Mn_3Si_2Te_6$ | This work | KF | 72.77, 72.75 | 0.42(9) | 1.18(5) | 3.75(9) |
| $1 \times 10^{16}$ ($H^+$/cm$^2$) $Mn_3Si_2Te_6$ | This work | KF | 73.29, 73.24 | 0.43(8) | 1.05(7) | 3.41(1) |
| $1 \times 10^{18}$ ($H^+$/cm$^2$) $Mn_3Si_2Te_6$ | This work | KF | 72.36, 72.28 | 0.44(6) | 1.06(5) | 3.38(9) |
| $Mn_3Si_2Te_6$ | [15] | KF | 74.18(8), 74.35(5) | 0.41(1) | 1.21(2) | 3.75(3) |
| $Mn_3Si_2Te_6$ | [14] | Neutron Diffraction | 78 | 0.25 | | |
| Mean Field Model | [27, 30] | Theory | | 0.5 | 1.0 | 3.0 |
| 3D Heisenberg Model | [27] | Theory | | 0.365 | 1.386 | 4.8 |
| 3D Ising Model | [27] | Theory | | 0.325 | 1.241 | 4.82 |
| 3D XY Model | [27] | Theory | | 0.345 | 1.316 | 4.81 |

Critical exponents can be determined by the KF method following [37],

$$\frac{M_S(T)}{dM_S(T)/dT} = \frac{T - T_C}{\beta}, \qquad (6)$$

$$\frac{\chi_0^{-1}(T)}{d\chi_0^{-1}(T)/dT} = \frac{T - T_C}{\gamma}, \quad . \qquad (7)$$

Figure 2(d) shows the KF plot represented with linear fits following Eqs. (6) and (7). The Widom scaling law, $\delta = 1 + \gamma/\beta$ [39], is used to calculate $\delta$ with the critical exponents obtained from the KF method. Table I shows the values for the critical parameters for all fluences in this work with reference to other critical exponent analyses and theoretical quantities. MST roughly exhibits an increasingly mean field-like behavior as a function of proton irradiation where $\beta$ increased and $\gamma$ decreased from their pristine counterpart. However, an anomaly appears at $5 \times 10^{15}$ $H^+$/cm$^2$ in which a sharp increase in $\gamma$ and slight decrease in $\beta$ was observed in comparison to pristine and other fluences, respectively (see Supplemental Material Figure S6 (a) and (b) [34]). Moreover, long-range interactions are presumed to exist in each sample of MST as the critical exponents $\beta$ and $\gamma$ lay closer to mean field critical exponents relative to the other universality classes presented in Table I.

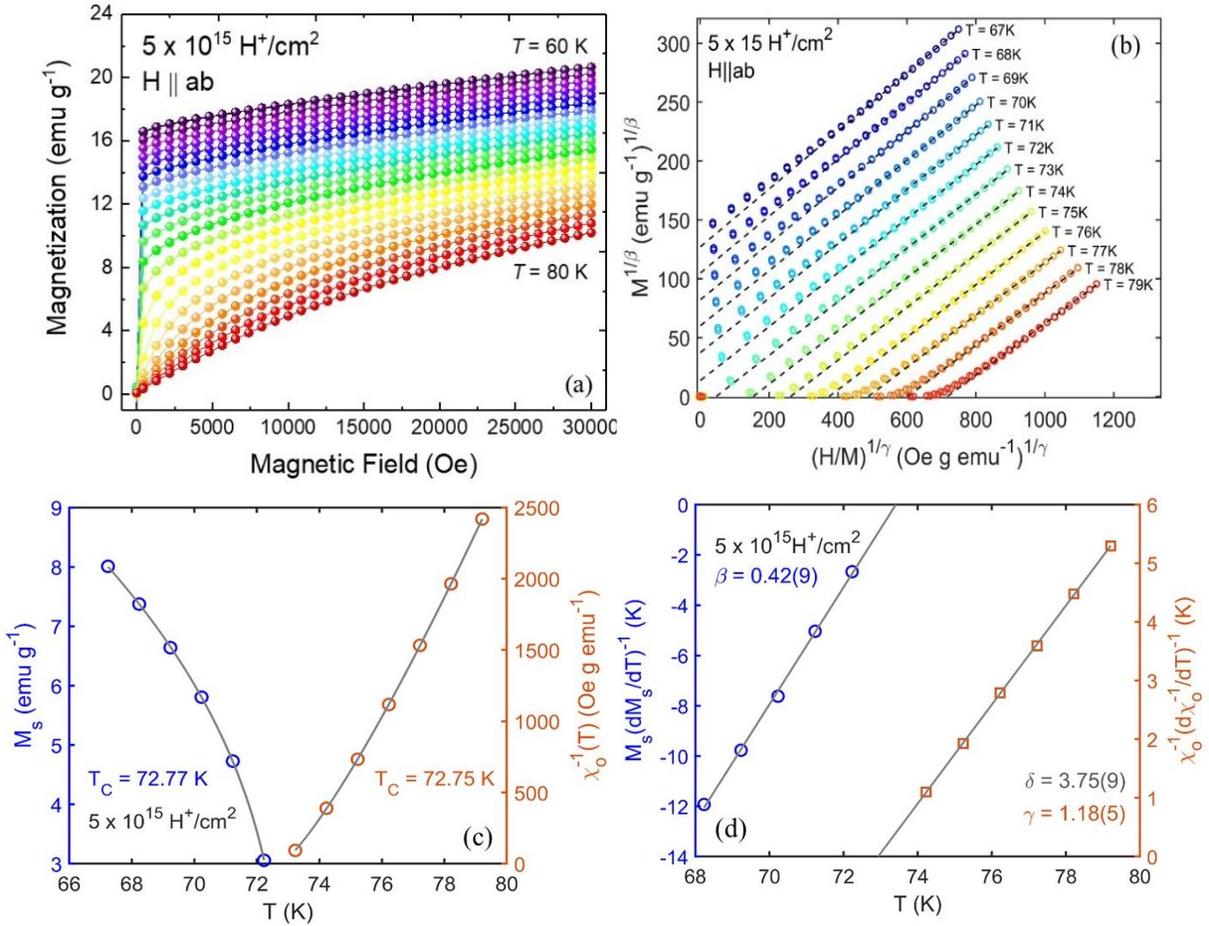

FIG. 2. (a) 1st quadrant isothermal magnetization curves from 60 to 80 K for H $\parallel$ $ab$ 5$\times$ $10^{15}$ H$^+$/cm$^2$ MST. (b) MAP for 5$\times$ $10^{15}$ H$^+$/cm$^2$ MST with a linear fit on isotherms with KF critical exponents. (c) Spontaneous magnetization and initial inverse susceptibility analysis displaying corresponding $T_C$'s. (d) Kouvel-Fisher plot for 5$\times$ $10^{15}$ H$^+$/cm$^2$ with critical exponents beta, gamma, and delta using the Widom scaling relation.

The reliability of the critical exponents and $T_C$ obtained from KF and MAP methods are confirmed by following Eq. (4) plotted as scaled $m$ vs $h$. Figure 3(a) shows the data collapsing into two branches, one above and below $T_C$ for the sample irradiated at 5 $\times$ $10^{15}$ H$^+$/cm$^2$. The two branched curves were present in all pristine and irradiated samples indicating that the correct critical regime was used for obtaining the critical exponents. The inset of Figure 3(a) shows the log-log scale of scaled $m$ and $h$ for 5 $\times$ $10^{15}$ H$^+$/cm$^2$. A more rigorous confirmation of the critical exponents was done by plotting scaled $m^2$ vs $h/m$ where two independent branches of data further confirm the reliability of the obtained exponents. This was observed in all pristine and irradiated samples shown in Supplemental Material [34] and is shown in Figure 3(b) for 5 $\times$ $10^{15}$ H$^+$/cm$^2$. The scaling equation of state can take on another form where all the curves should collapse into a single curve:

$$\frac{H}{M^\delta} = k\left(\frac{\varepsilon}{H^{1/\beta}}\right), \qquad (8)$$

where $k(\varepsilon, H)$ is the scaling function. The inset of Figure 3(b) shows $MH^{-1/\delta}$ vs $H^{-(\delta\beta)}$ with a single curve and the zero point of the horizontal axis locates the $T_C$, confirming the $T_C$ and critical exponents found for all proton irradiated samples of MST. The scaling equation of state is clearly satisfied in the critical regime indicating that the interactions have been properly renormalized. See Supplemental Material Figures S2–S5 for all plots with converged results for the critical behavior analysis and scaling relations [34].

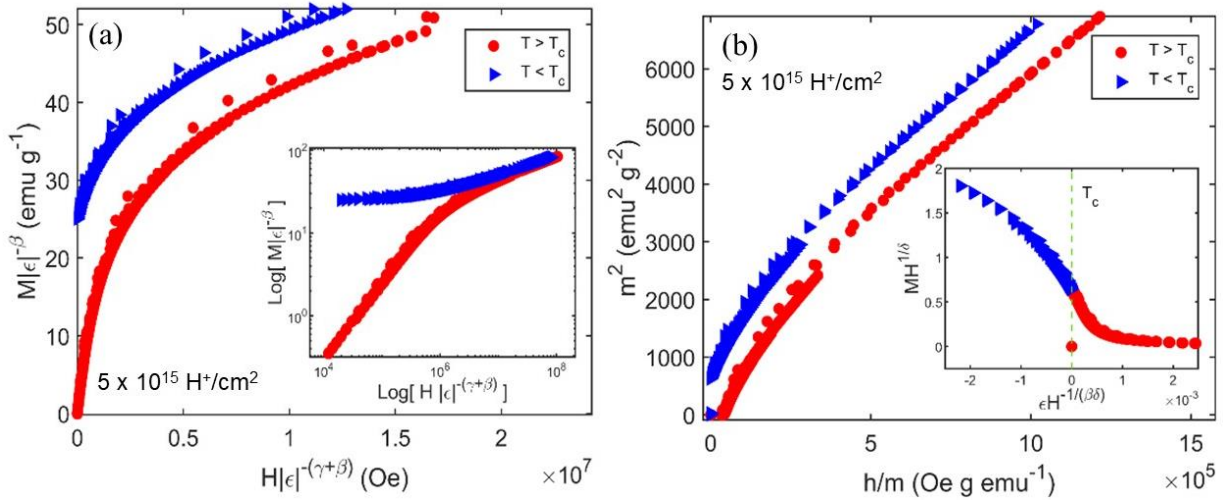

FIG. 3. (a) Scaling of renormalized magnetization, $m$, and renormalized field, $h$, with isotherms falling into two branches, below and above $T_C$ for $5\times 10^{15}$ H$^+$/cm$^2$ MST. Inset: The log-log scale of the renormalized $m$ and $h$ for the pristine case. (b) Data from $5\times 10^{15}$ H$^+$/cm$^2$ MST where isotherms collapse into two separate branches, one below and one above $T_C$, further confirming the reliability of the exponents in the critical regime. Inset: Experimental data for $5\times 10^{15}$ H$^+$/cm$^2$ MST collapsing into one single curve where $T_C$ is located at the zero point of the horizontal axis.

It is important to determine the range of interaction in proton irradiated MST to understand how it is affected by proton bombardment. The universality class of the magnetic phase transition for a homogeneous magnet with isotropic long-range interactions depends on the exchange distance $J(r) = r^{-(d+\sigma)}$, where $r$ is the distance and $\sigma$ is the range of interaction [40]. It is worth noting that competing exchange and superexchange interactions disqualify MST from being categorized as an "isotropic magnet". However, as will be shown in the results, the long-range nature of the magnetic interactions makes this model a fair approximation to suitably study trends in the spin correlations. From this model, the spin interaction can either be long or short depending on whether $\sigma < 2$ or $\sigma > 2$, respectively. According to renormalization group theory, the susceptibility exponent $\gamma$ can be calculated as,

$$\gamma = 1 + \frac{4}{d}\left(\frac{n+2}{n+8}\right)\Delta\sigma + \frac{8(n+2)(n-4)}{d^2(n+8)^2} \times \left[1 + \frac{2G\left(\frac{d}{2}\right)(7n+20)}{(n-4)(n+8)}\right]\Delta\sigma^2, \qquad (9)$$

where $\Delta\sigma = \sigma - \frac{d}{2}$ and $G\left(\frac{d}{2}\right) = 3 - \frac{1}{4}\left(\frac{d}{2}\right)^2$, $d$ is the effective spatial dimensionality and $n$ is the spin dimensionality [41]. This method is carried out by inserting experimentally gained $\gamma$ into Eq. (9) for different values of $d$ and $n$. Values of $d$ and $n$ are confirmed by using additional critical exponent expressions that should match well with experimentally found exponents. These remaining exponents are obtained from the following expressions: $\eta = 2 - \sigma$, $\nu = \gamma/\sigma$, and the following scaling relations: $\alpha = 2 - \nu d$, $\beta = (2 - \alpha - \gamma)/2$.

For pristine MST, we find that the $J(r)$ decreases as $r^{-4.71}$. This result is close to the previously reported exchange distance for pristine MST, as seen in Table II [13]. For pristine MST, we obtain $d = 3$ and $n = 1$ with a $\sigma = 1.71$. Moreover, the effective spatial dimensionality remains as $d = 3$ for all samples (see Table II). The Ising spin dimensionality ($n = 1$) indicates anisotropic magnetic interactions and is present for the pristine case, along with the samples irradiated with $1 \times 10^{16}$, and $1 \times 10^{18}$ H$^+$/cm$^2$ [42]. Interestingly, the spin dimensionality increases to $n = 2$ and $n = 3$ for $1 \times 10^{15}$ and $5 \times 10^{15}$ H$^+$/cm$^2$, indicating an $XY$ and a Heisenberg interaction, respectively. This observation of the change in spin dimensionality at $1 \times 10^{15}$ and $5 \times 10^{15}$ H$^+$/cm$^2$ is intriguing as a change of this type would indicate a crossover in the spin model and a change in the magnetic exchange interaction strength. As suggested by Fisher in Ref. [42], frustration or spatial anisotropy in Heisenberg spins can induce a crossover to $XY$ or Ising spins. It is known that MST exhibits magnetic frustrations which arise from competing ferromagnetic superexchange interactions between Mn-Te-Mn bonds and the anisotropic direct exchange interactions between Mn sites [12]. Therefore, one may safely assume that the origin of this crossover behavior is due to magnetic frustrations present in the MST compound. The root of this variation should be investigated further to quantitatively understand the effects proton irradiation has on the spin dimensionality and the magnetic state of MST. Employing an anisotropic critical behavior study should also reveal significant changes in the magnetic anisotropy for magnetization measurements performed for in plane and out of plane magnetic fields. In addition, spin-polarized density functional theory calculations can reveal the mechanism in which proton irradiation is influencing the magnetic interactions [43].

Table II. Exchange distance decay function J(r) and critical exponents for each MST fluence. Here, σ is the range of interaction and the spatial dimensionality is expressed as d. Additionally, spin dimensionality is expressed as n, and ν as the correlation length. η describes the variation of the spin-spin correlation function with distance at $T_C$. These critical exponents are calculated using our experimentally found γ from our KF method analysis.

| Fluence (H$^+$/cm$^2$) | J(r) | σ | d | n | ν | η |
|---|---|---|---|---|---|---|
| Pristine | r$^{-4.71}$ | 1.71 | 3 | 1 | 0.649 | 0.29 |
| $1 \times 10^{15}$ | r$^{-4.62}$ | 1.616 | 3 | 2 | 0.661 | 0.384 |
| $5 \times 10^{15}$ | r$^{-4.74}$ | 1.743 | 3 | 3 | 0.68 | 0.257 |
| $1 \times 10^{16}$ | r$^{-4.62}$ | 1.617 | 3 | 1 | 0.654 | 0.383 |
| $1 \times 10^{18}$ | r$^{-4.63}$ | 1.632 | 3 | 1 | 0.653 | 0.368 |

| Pristine [15] | $r^{-4.79}$ | 1.79 | 0.676 |

To gain a better understanding of the spin correlations and the critical phenomena, the critical exponent $\eta$ is calculated using $\eta = 2 - \frac{\gamma d}{2\beta + \gamma}$. The critical exponent is used to calculate the two-point correlation function $G(r) = 1/r^{-d+2-\eta}$, which describes how spins are related in a system. A decrease in the spin correlation decay as the proton irradiance is increased is observed. This indicates a decrease in magnetic ordering, except at the fluence of $5 \times 10^{15}$ H$^+$/cm$^2$. The sample irradiated with $5 \times 10^{15}$ H$^+$/cm$^2$ displayed the smallest correlation function decay rate, overall indicating a larger correlation of the spins. Interestingly, this trend observed for the correlation function decay rate (Fig. 4) was found to match our previous magnetic characterization study of proton irradiated Mn$_3$Si$_2$Te$_6$, where the sample irradiated with $5 \times 10^{15}$ H$^+$/cm$^2$ possesses the largest magnetization [14]. Additionally, Ref. [14] also shows magnetization decreasing for the samples at $1 \times 10^{15}$, $1 \times 10^{16}$, and $1 \times 10^{18}$ H$^+$/cm$^2$ relative to the pristine sample, in perfect agreement with the trends observed in the spin correlation decay rate in the present study (see Fig. 4). Therefore, the two-point correlation function, combined with our analysis of critical behavior, was found to be a valuable tool for characterizing the magnetic properties of MST following proton irradiation.

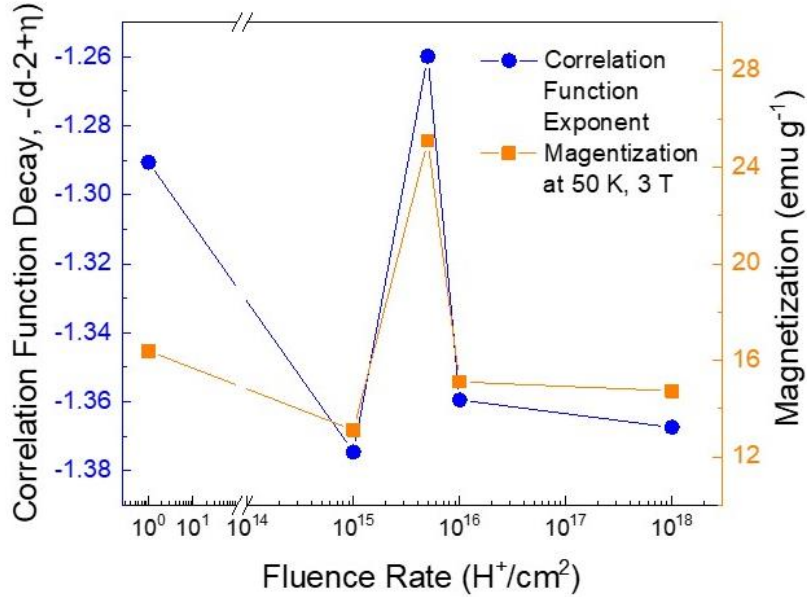

FIG. 4. Correlation function (left, blue) and 50 K magnetization taken at 3 T (right, orange) plotted as a function of proton irradiation. A maximum peak at $5 \times 10^{15}$ H$^+$/cm$^2$ is observed for both plots.

## B. Magnetic change in entropy from magnetization and heat capacity

The negative change in magnetic entropy, $-\Delta S_M(T)$, is evaluated to examine how it is fluctuating as a function of proton fluence. The greatest peak in $-\Delta S_M(T)$ will indicate the sample containing the largest increase in magnetic interactions (i.e., magnetization and spin correlations) as $-\Delta S_M(T)$ is associated with the spin configuration order of a material. Moreover, this is done to confirm the increase in magnetic interactions and spin fluctuations present at $5 \times 10^{15}$ H$^+$/cm$^2$. According to classical thermodynamic theory, isothermal magnetic entropy change with external magnetic fields is given by [44–46]:

$$\Delta S_M(T,H) = \int_0^H \left(\frac{\partial S(T,H)}{\partial H}\right)_T dH = \int_0^H \left(\frac{\partial M(T,H)}{\partial T}\right)_H dH, \quad (10)$$

using Maxwell's relation for magnetic entropy and magnetization by $\left(\frac{\partial S}{\partial H}\right)_T = \left(\frac{\partial M}{\partial T}\right)_H$, where $H$ and $T$ are applied magnetic fields and constant temperatures, respectively. For discrete magnetic field and temperature intervals of magnetization measurements, $\Delta S_M$ is approximated as,

$$\Delta S_M(T,H) = \frac{\int_0^H M(T_{i+1},H)dH - \int_0^H M(T_i,H)dH}{T_{i+1}-T_i}, \quad (11)$$

where $M(T_{i+1},H)$ and $M(T_i,H)$ are magnetization at $T_{i+1}$ and $T_i$ under applied magnetic fields, respectively. Pristine and proton irradiated plots of the $-\Delta S_M$ at select fields from 1 to 3 T are shown in Supplemental Material Figure S7 [34]. The evolution of maximum $-\Delta S_M$ as a function of proton irradiation for magnetic fields applied in the *ab*-plane are presented in Figure 5(a). The region in which the change in magnetic entropy is maximum, $-\Delta S_M^{max}$, indicates the $T_C$. A clear increase in the absolute value of $-\Delta S_M$ with increasing magnetic fields from 1 to 3 T is observed and shown in Supplemental Material Figure S7 [34]. The $-\Delta S_M^{max}$ at 3 T is the largest for $5 \times 10^{15}$ H+/cm$^2$ with a value of 2.45 J/kg K. The corresponding temperatures ($\sim T_C$) in the ferromagnetic transition and the $-\Delta S_M$ peaks for each sample are seen in Table III. Overall, the largest $-\Delta S_M^{max}$ is seen for $5 \times 10^{15}$ H+/cm$^2$ in Figure 5(a). Likewise, the smallest spin is in parallel to the maximums in the spin correlation decay rate and largest magnetization (50 K, 3 T) also occurred at $5 \times 10^{15}$ H+/cm$^2$ (Fig. 4) [14]. According to Maxwell's relation, magnetic entropy change is directly proportional to change in magnetization; therefore, these findings further confirm that proton irradiation performed at $5 \times 10^{15}$ H+/cm$^2$ distinctly increased the spin correlations and enhanced the magnetization. This further corroborates the idea that there is a strengthened ferromagnetic interaction occurring at the fluence rate of $5 \times 10^{15}$ H+/cm$^2$, which produces enhanced magnetization.

Table III. Maximum change in magnetic entropy ($\Delta S_M^{max}$) for MST irradiated samples at 3 T for H ∥ *ab*. Corresponding peak temperature ($T_C$) at each maximum is also shown.

| Fluence (H+/cm$^2$) | $\Delta S_M^{max}$ (J/kg K) | $T_C$ (K) |
|---|---|---|
| Pristine | 1.60 | 72.99 |
| $1 \times 10^{15}$ | 1.26 | 73.97 |
| $5 \times 10^{15}$ | 2.45 | 73.66 |
| $1 \times 10^{16}$ | 1.52 | 73.60 |
| $1 \times 10^{18}$ | 1.60 | 73.04 |

The thermodynamic nature of the phase transition and the transition temperature are examined by performing heat capacity measurements with no magnetic field in the high-temperature regime spanning 50 – 300 K. The temperature dependence of heat capacity is shown in Figure 5(b) for all MST samples. A peak appears at 72.23, 72.06, 72.11, 71.83, 71.72 K for pristine, $1 \times 10^{15}$, $5 \times 10^{15}$, $1 \times 10^{16}$, and $1 \times 10^{18}$ H$^+$/cm$^2$, respectively. The local maximum(s) in the data confirm that the ferromagnetic transition occurs in the vicinity of ~72 K and clearly indicate a bulk PM to FM transition in the MST compounds, in agreement with isothermal magnetization and critical behavior analysis.

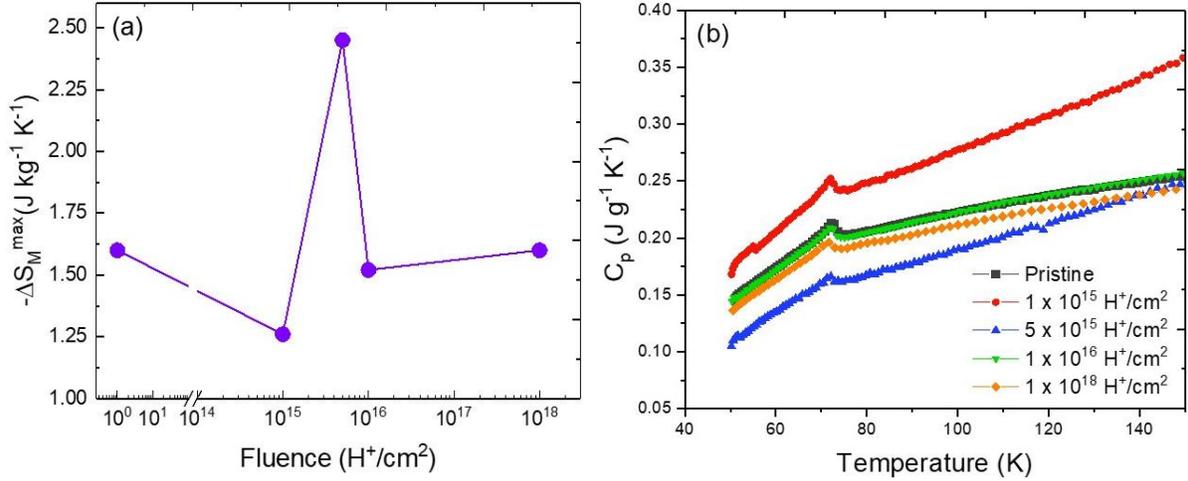

FIG. 5. (a) Maximum magnetic entropy change as a function of proton fluence, where $5 \times 10^{15}$ H$^+$/cm$^2$ has largest change. (b) Heat capacity as a function of temperature plotted for all proton irradiated samples.

To extrapolate the magnetic entropy released across the magnetic transition ($S_M$), the method described by May *et. al.* is employed which considers the lack of suitable phonon reference material and a negligible electron contribution from heat capacity data [12]. Heat capacity is initially plotted versus temperature ($C_p/T$), then the range from 60 to 95 K is isolated as shown in Supplemental Material Figure S8 [34]. In this region, a linear background is subtracted to account for the previous considerations. Additionally, this is done to intentionally over-estimate the magnetic entropy in the system to fully capture magnetic contributions to the data. An explanation for this overestimation is discussed later. The $S_M$ is then estimated from heat capacity measurements by integrating over the isolated interval,

$$S_M = \int_0^T \frac{C_M}{T} dT. \qquad (12)$$

Upon repeating the procedure for all proton irradiated samples, magnetic entropy derived from heat capacity does not show large deviations across the range of samples, as seen in Fig. 6(a). The point at which the magnetic entropy saturates is taken as the maximum entropy released across the transition. Small variations are attributed to noise and differences in the linear backgrounds.

Nevertheless, the analysis yields very small numbers even after a deliberate over-estimation. See Supplemental Material Table S1 for values of $S_M$ [34].

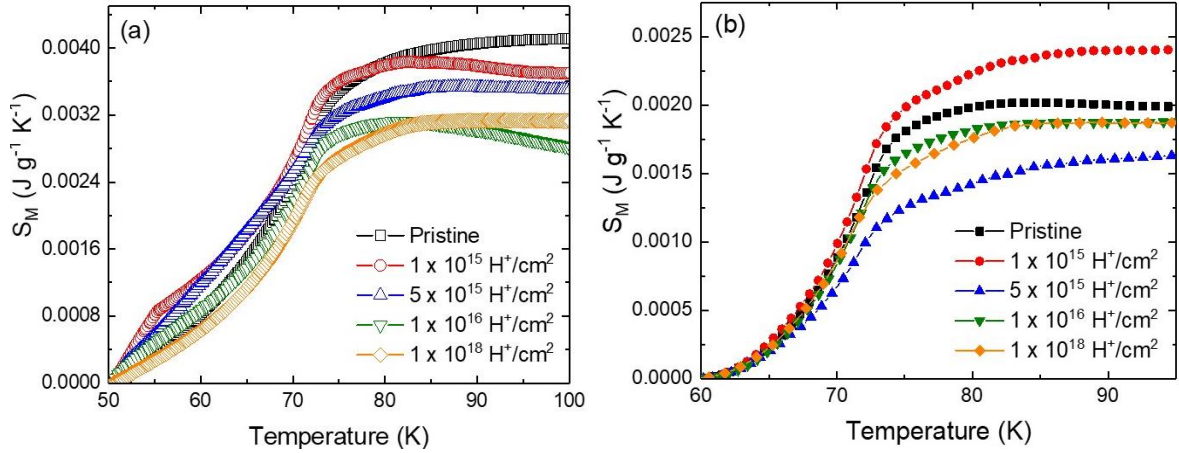

FIG. 6. (a) Magnetic entropy estimated from heat capacity analysis utilizing a linear background subtraction of the heat capacity plot (b) Magnetic entropy estimated from heat capacity analysis utilizing Einstein fits to isolate the magnetic contribution by subtracting the lattice contribution.

A similar analysis utilizing Einstein fits reinforces the conclusion taken from the previous analysis. For this analysis, the data were fitted using the sum of two Einstein fits of the form:

$$C_E = \sum 3Rn_i \left(\frac{\theta_i}{T}\right)^2 \frac{e^{\theta_i/T}}{\left(e^{\theta_i/T} - 1\right)^2}. \tag{13}$$

In this model, $R$ denotes the specific gas constant for MST, $n_i$ is a weight factor, and $\theta_i$ is the Einstein temperature. Additional terms for a more accurate fit would include a Debye term, however, necessary data in the temperature range of $2 - 50$ K is unavailable at the moment. Other considerations include more reference materials and an appropriate electron contribution. Despite the lack of these, two Einstein fits are enough for the purpose of this analysis. For these fits, only the high temperature data above the magnetic transition is selected and vary slightly across the samples to provide an optimal fit. The fitting curves act as a simulated lattice contribution and are subtracted from the total heat capacity to isolate the magnetic contribution to the data, see Supplemental Material Figure S9 [34]. As in the previous analysis, this is an intentional overestimation and variations are attributed to differences in fitting parameters [47]. This method also shows that the magnetic entropy is comparable across fluences from heat capacity analysis as shown in Supplemental Material Table S2 [34]. Although this method yields higher values of S$_M$, they are still drastically smaller than those expected from $R \ln(2S + 1)$, which yields magnetic entropy change at the ordering temperature. Furthermore, an even larger overestimation still yields small values of released magnetic entropy.

In MST, the magnetic entropy per mole of Mn in its paramagnetic state for Mn$^{2+}$ ($S = 5/2$) is given by $S = R \ln(6) = 0.015$ J/K. However, in comparison to the two values given by the previous analyses (see Fig. 4) the S$_M$ values are an order of magnitude smaller. This behavior is typical of materials with frustrated short-range correlations or competing interactions [47,48], and are attributed to the magnetic frustration caused by the anti-ferromagnetic (AFM) coupling between the Mn1 and Mn2 sites. Similar values in magnetic entropy are observed in the layered AFM MnPS$_3$, where 2D order is observed above $T_C$ [49,50] as well as in the layered ferromagnet CrSiTe$_3$, where short-range correlations are observed above $T_C$ and the magnetism is strongly coupled to the lattice [51,52]. Such small magnetic entropy values were reported by May *et. al.* on MST for easy axis measurements [12]. Despite the small variations in S$_M$ across all fluences, there are no considerable changes in released magnetic entropy after proton irradiation, suggesting a stable AFM coupling between the Mn sites. In other words, proton irradiation does not seem to weaken or strengthen the AFM coupling, implying that it is not the manner in which the magnetization is modified in Ref. [14]. In fact, this may support the theory that the magnetization is enhanced through a strengthening of the super-exchange interaction between Mn atoms mediated through Te rather than a weakening of the AFM component.

## IV. CONCLUSIONS

We have identified the critical exponents for proton irradiated Mn$_3$Si$_2$Te$_6$ while examining the spin and spatial dimensionality, spin correlations, and magnetic entropy. We found that increasing the proton fluence leads to an increasingly mean field-like behavior, however the samples ultimately did not fall into one specific universality class. Exchange distance analysis showed that the decay rate of magnetic interaction is the smallest for $5 \times 10^{15}$ H$^+$/cm$^2$ compared to pristine and other irradiated samples. The spin correlations and maximum change in magnetic entropy at 3 T indicate an increase in magnetic interactions for $5 \times 10^{15}$ H$^+$/cm$^2$, in line with the increase in magnetization as seen in Martinez *et al.* [14]. It has been previously established that frustration in MST is caused by the short-range interactions of the AFM coupled Mn atoms [12]. Heat capacity analysis confirms the presence of frustration in all MST samples demonstrated by the low amounts of magnetic entropy released even below $T_C$. Therefore, the in-plane frustration in MST is thought to be the cause of the crossover critical phenomena. This crossover behavior is manifested differently for each sample, as observed in this study where $n = 1$ for pristine MST transforms to $n = 2$ and $n = 3$ for $1 \times 10^{15}$ and $5 \times 10^{15}$ H$^+$/cm$^2$, respectively. Additionally, the fact that AFM coupling is not diminished even above $Tc$ is worth investigating further to explore the possibility of an additional AFM transition at higher temperatures. Further studies on the critical phenomena on MST can be carried out in the fluence range spanning $10^{15}$ H$^+$/cm$^2$ to illuminate the magnetic behavior in this region. In short, examination of the critical phenomena can be extended to other systems that involve external perturbations of vdW materials and/or their few and mono-layer counterparts.

## ACKNOWLEDGMENTS

This work was prepared by R. Olmos, J. A. Delgado, H. Iturriaga, L. M. Martinez, C. L. Saiz, and S. R. Singamaneni, and co-authors under the award number 31310018M0019 from The University of Texas at El Paso (UTEP), Nuclear Regulatory Commission. The statements, findings, conclusions, and recommendations are those of the author(s) and do not necessarily reflect the view of UTEP or The Nuclear Regulatory Commission. Authors also acknowledge the UTEP start-up fund in supporting this work. Work at Brookhaven National Laboratory is supported by the

## SUPPLEMENTAL MATERIAL

### A. Isothermal Magnetization

In order to begin magnetic critical behavior analysis we perform isothermal magnetization measurements (M-H) reserved to the first quadrant. Each temperature isotherm from the first quadrant M-H will be used for magnetic critical behavior and magnetic entropy change analysis. Figure S1 (a) and (b) confirm the easy axis and hard axis of the pristine (unirradiated) $Mn_3Si_2Te_6$ (MST) sample through isothermal magnetization for H ∥ $ab$ and H ∥ $c$, respectively. Figure S1 (c)-(e) show isothermal magnetization for the other proton irradiated MST specimens examined in this study. The isotherms in Figure S1 are between the ranges of 61 to 81 K, near the critical regime of MST whose Curie temperature ($T_C$) lies in the range between $74 - 78$ K [1–3].

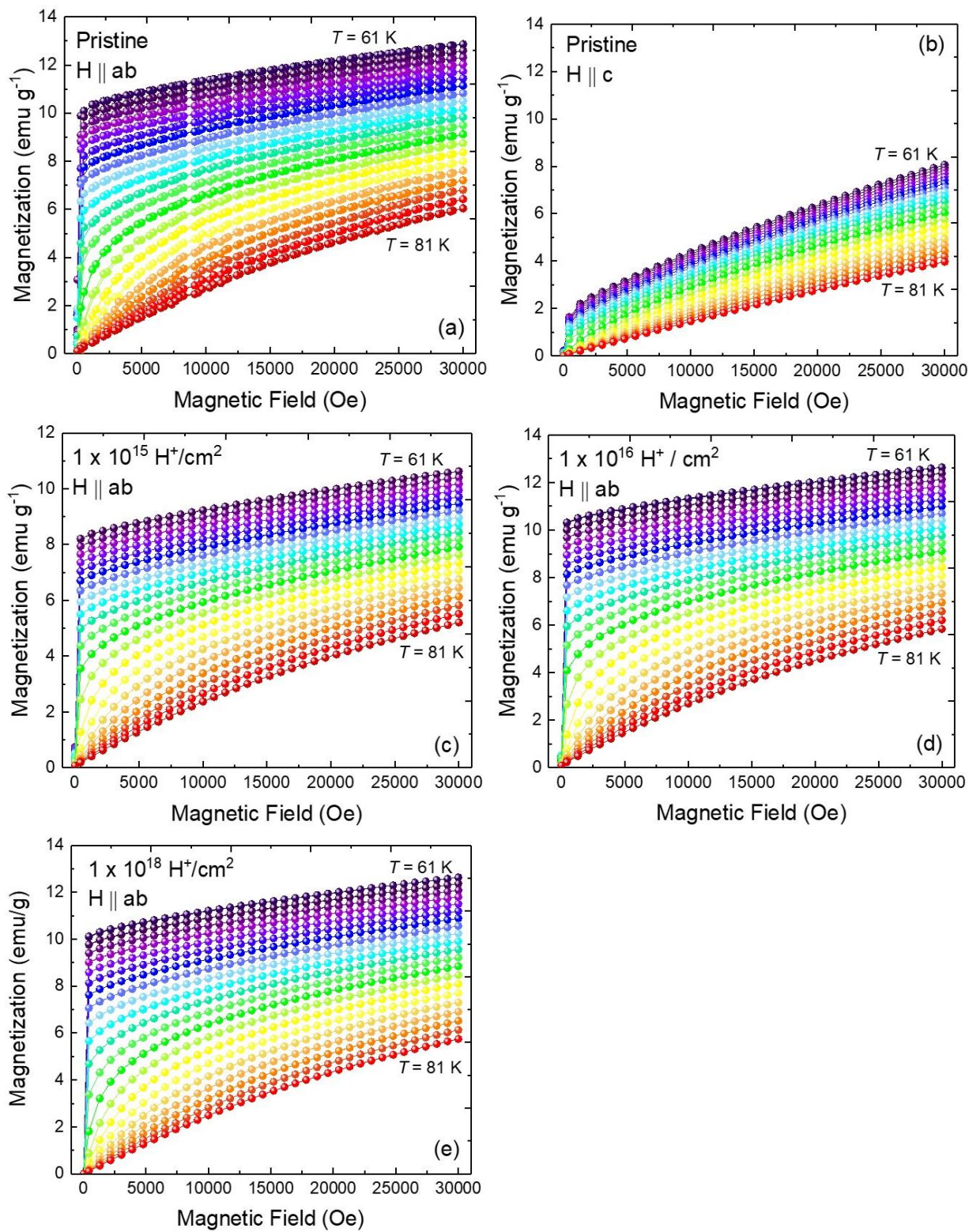

FIG. S1. First quadrant isothermal magnetization curves for pristine (a) H ∥ *ab* and (b) H ∥ *c* and proton irradiated samples of MST at (c) 1 x 10^15, (d) 1 x 10^16, and (e) 1 x 10^18 H+/cm². Isotherms are shown decreasing as the temperature increases from 61 to 81 K. Magnetization measurements were conducted in the easy axis where the magnetic field, H, is parallel to the *ab* plane for the proton irradiated samples.

### B.  Pristine (H ‖ ab) $Mn_3Si_2Te_6$ Critical Analysis and Scaling Relations

Figure S2 displays the initial critical behavior we first performed for MST on the pristine crystal for magnetic field aligned with the *ab* crystal plane. Figure S2 (a) shows the Arrott plot ($M^2$ v. $H/M$) with isotherms in the critical regime. The high field region is described relatively well with a linear fit but not quite satisfied. Therefore, we move onto performing an iterative critical analysis using Kouvel-Fisher (KF) and modified Arrott plot (MAP) methods for pristine and proton irradiated samples of MST. Figure S2 (b) shows the converged plot for pristine MST after applying the KF method giving the final critical exponents $\beta$, $\gamma$, and calculated $\delta$ using the Widom scaling law. Similarly, Figure S2 (c) shows the final power law fits of the temperature dependent spontaneous magnetization ($M_S$) and inverse initial susceptibility ($\chi_0^{-1}$) where the intersection with the x-axis indicates the magnetic order transition temperature. Figure S2 (d) shows the linear fitting of the MAP after the iterative process converged. It is clear that even the low field region linear fitting has improved. We perform several scaling relations to confirm the reliability of our critical behavior analysis within the critical regime we used. Figure S2 (e) displays the theoretical $\delta$ exponent using the power law relation $M = D \ H^{1/\delta}$ where $T = T_C$; a linear fit of the log-log scale of $M$ vs. $H$ gives the critical exponent $\delta$. Figure S2 (f) and (g) display another form of the scaling equation of state in two ways; $M$ and $H$ in terms of reduced temperature ($\varepsilon$) and the scaling function $k(x)$; and the scaled magnetization, $m$, and scaled magnetic field, $h$, respectively. For both of these plots, the numbers should fall into two sections of data, one below and above $T_C$. If this is observed, one can be sure that they took the correct critical regime into play for analysis. In these figures we observe such branching of the data into two separate sections. Further reliability is seen in the inset of Figure S2 (f), which shows the log-log scale of the $M$ and $H$ in terms of $\varepsilon$ where two branches fall into one, an indication that the critical regime used satisfies the scaling relations. Furthermore, inset of Figure S2 (g) has the plot $MH^{-1/\delta}$ vs. $\varepsilon H^{-1/(\beta\delta)}$, where the data should collapse into a singular curve. The zero point of the horizontal axis also indicates the $T_C$.

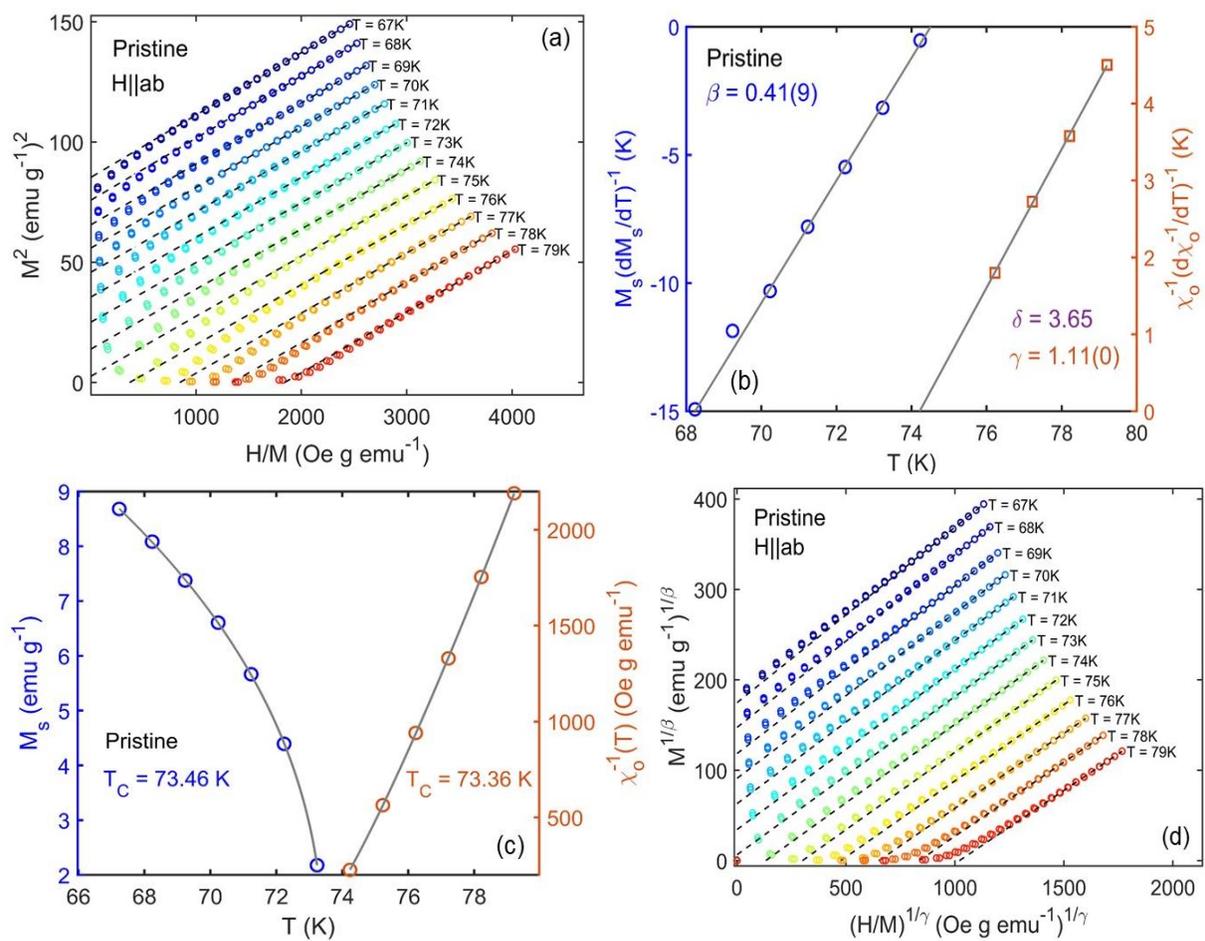

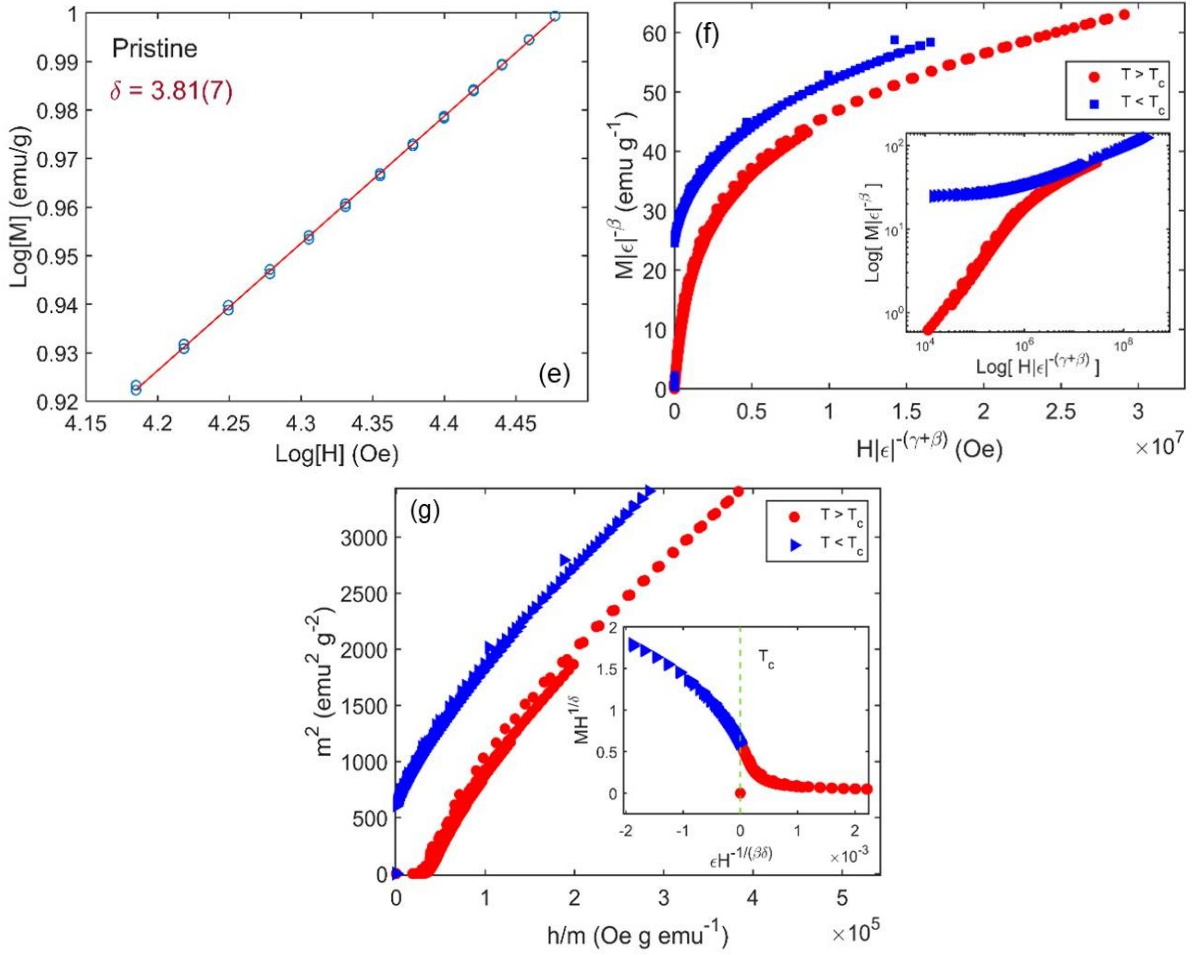

FIG. S2. (a) Arrott plot for pristine MST displaying a large mean field behavior, however, there is still a curvature in the low field region leading to a more in depth critical analysis for this compound. (b) Kouvel-Fisher plot for pristine MST with critical exponents $\beta$, $\gamma$, and $\delta$ using the Widom scaling relation. (c) Magnetization and susceptibility analysis. (d) MAP for pristine MST with a linear fit on isotherms with KF critical exponents. (e) Critical isotherm plot obtaining critical exponent $\delta$. (f) Scaling of renormalized magnetization, $m$, and renormalized field, $h$, with isotherms falling into two branches, below and above $T_C$ pristine MST. Inset shows the log-log scale of the renormalized m and h for the pristine case. (g) Data from pristine H ∥ $ab$ MST isotherms collapse into two separate branches, one below and one above $T_C$, which further confirm the reliability of the exponents in the critical regime. Inset shows experimental data for pristine MST collapsing into one single curve where $T_C$ is located at the zero point of the horizontal axis.

## C. $1 \times 15$ H$^+$/cm$^2$ (H ∥ ab) MST Critical Analysis and Scaling Relations

Figure S3 (a), (b), and (c) show the converged plots for the KF method with critical exponents $\beta$ and $\gamma$, the $M_S$ and $\chi_0^{-1}$ power law fits with respective $T_C$'s, and the MAP with a linear fit showing the best fit for this critical regime, respectively for the proton irradiation at $1 \times 10^{15}$ H$^+$/cm$^2$. Figure S3 (d) and (e) show the scaling relations $M|\varepsilon|^{-\beta}$ vs. $H|\varepsilon|^{-(\gamma+\beta)}$ and $m^2$ vs. $h/m$, respectively where two branches of data should form, one below and one above the $T_C$. The insets of Figure S3 (d) and (e) show the log $(M|\varepsilon|^{-\beta})$ vs. log $(H|\varepsilon|^{-(\gamma+\beta)})$ where two branches should fall into one, and the $MH^{-1/\delta}$ vs. $\varepsilon H^{-1/(\beta\delta)}$ where the zero point of the horizontal axis indicates the $T_C$, respectively. From these scaling relation plots we confirm that the magnetic critical behavior analysis we performed for the $1 \times 10^{15}$ H$^+$/cm$^2$ MST sample is true and properly scaled, thus yielding reliable critical exponents $\beta$, $\gamma$ and $\delta$.

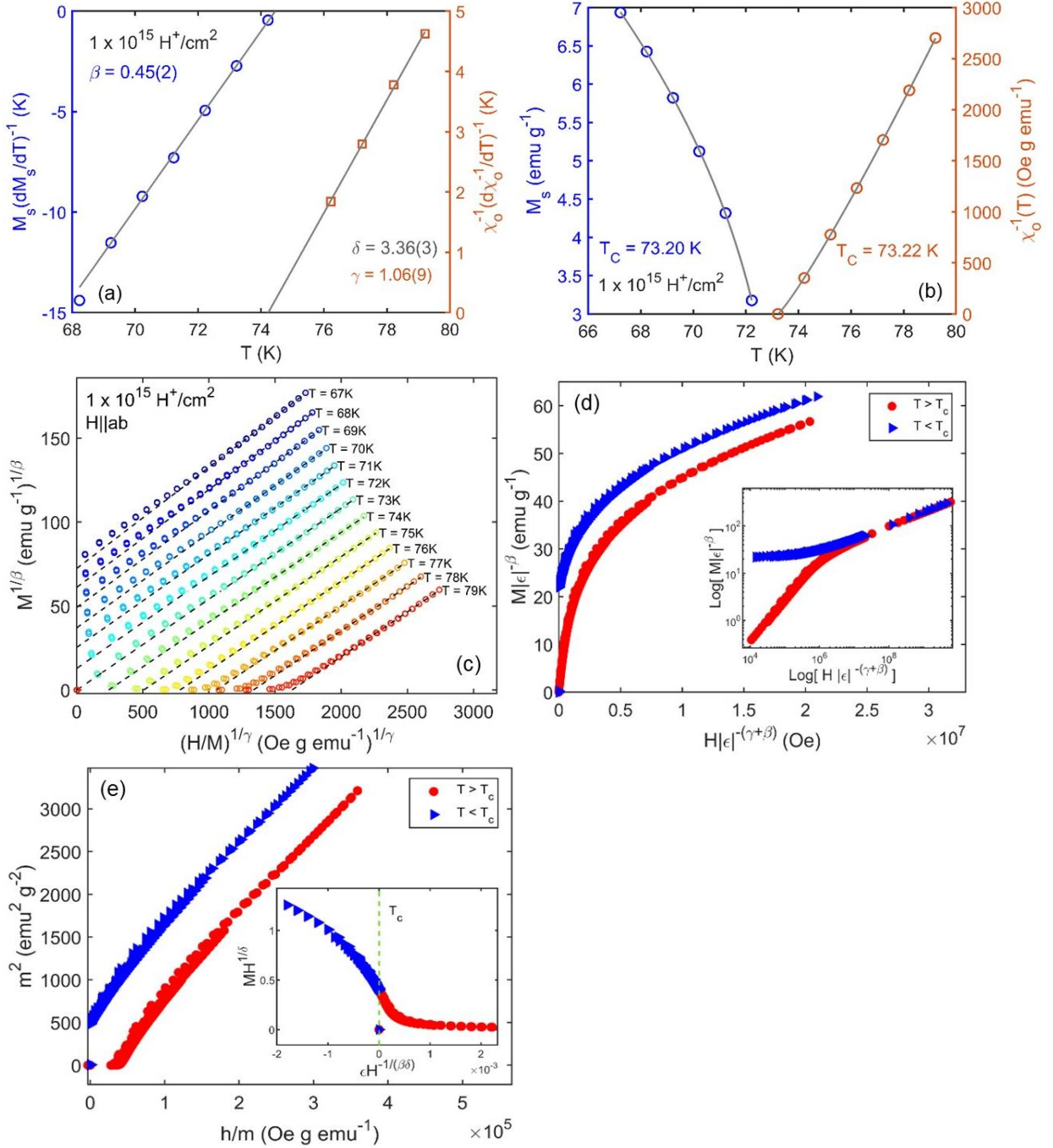

FIG. S3. (a) Kouvel-Fisher plot for $1 \times 15$ H$^+$/cm$^2$ Mn$_3$Si$_2$Te$_6$ with critical exponents $\beta$, $\gamma$, and $\delta$ using the Widom scaling relation. (b) Magnetization and susceptibility analysis. (c) MAP for $1 \times 15$ H$^+$/cm$^2$ Mn$_3$Si$_2$Te$_6$ with a linear fit on isotherms with KF critical exponents. (*d*) Scaling of renormalized

magnetization, $m$, and renormalized field, $h$, with isotherms falling into two branches, below and above $T_C$ $1 \times 15$ H$^+$/cm$^2$ Mn$_3$Si$_2$Te$_6$. Inset shows the log-log scale of the renormalized $m$ and $h$ for the $1 \times 15$ H$^+$/cm$^2$ case. (e) Data from $1 \times 15$ H$^+$/cm$^2$ H $\parallel ab$ Mn$_3$Si$_2$Te$_6$ isotherms collapse into two separate branches, one below and one above $T_C$, which further confirm the reliability of the exponents in the critical regime. Inset shows experimental data for $1 \times 15$ H$^+$/cm$^2$ Mn$_3$Si$_2$Te$_6$ collapsing into one single curve where $T_C$ is located at the zero point of the horizontal axis.

### D. $1 \times 16$ H$^+$/cm$^2$ (H ∥ ab) Mn$_3$Si$_2$Te$_6$ Critical Analysis and Scaling Relations

Figure S4 (a), (b), and (c) show the converged plots for the KF method with critical exponents $\beta$ and $\gamma$, the $M_S$ and $\chi_0^{-1}$ power law fits with respective $T_C$'s, and the MAP with a linear fit showing the best fit for this critical regime, respectively for the proton irradiation at $1 \times 10^{16}$ H$^+$/cm$^2$. Figure S3 (d) and (e) show the scaling relations $M|\varepsilon|^{-\beta}$ vs. $H|\varepsilon|^{-(\gamma+\beta)}$ and $m^2$ vs. $h/m$, respectively where two branches of data should form, one below and one above the $T_C$. The insets of Figure S4 (d) and (e) show the $\log(M|\varepsilon|^{-\beta})$ vs. $\log(H|\varepsilon|^{-(\gamma+\beta)})$ where two branches should fall into one, and the $MH^{-1/\delta}$ vs. $\varepsilon H^{-1/(\beta\delta)}$ where the zero point of the horizontal axis indicates the $T_C$, respectively. From these scaling relation plots we confirm that the magnetic critical behavior analysis we performed for the $1 \times 10^{16}$ H$^+$/cm$^2$ MST sample is true and properly scaled, thus yielding reliable critical exponents $\beta$, $\gamma$ and $\delta$.

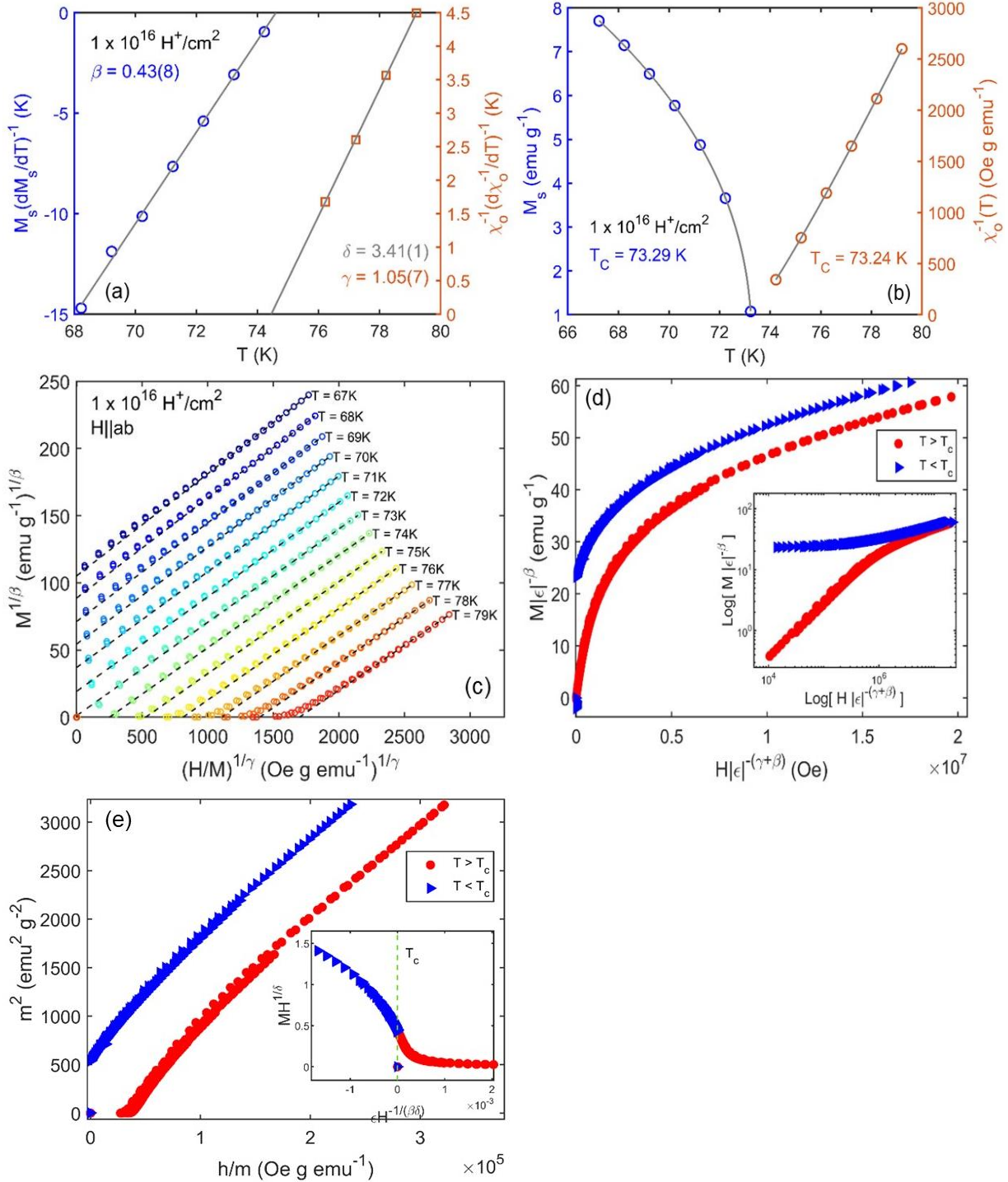

FIG. S4. (a) Kouvel-Fisher plot for $1 \times 10^{16}$ H$^+$/cm$^2$ Mn$_3$Si$_2$Te$_6$ with critical exponents $\beta$, $\gamma$, and $\delta$ using the Widom scaling relation. (b) Magnetization and susceptibility analysis. (c) MAP for $1 \times 10^{16}$ H$^+$/cm$^2$ Mn$_3$Si$_2$Te$_6$ with a linear fit on isotherms with KF critical exponents. (d) Scaling of renormalized magnetization, $m$, and renormalized field, $h$, with isotherms falling into two branches, below and above $T_C$ $1 \times 10^{16}$ H$^+$/cm$^2$ Mn$_3$Si$_2$Te$_6$. Inset shows the log-log scale of the renormalized $m$ and $h$ for the $1 \times 10^{16}$ H$^+$/cm$^2$ case. (e) Data from $1 \times 10^{16}$ H$^+$/cm$^2$ H $\parallel$ $ab$ Mn$_3$Si$_2$Te$_6$ isotherms collapse into two separate branches,

one below and one above $T_C$, which further confirm the reliability of the exponents in the critical regime. Inset shows experimental data for $1 \times 10^{16}$ H$^+$/cm$^2$ Mn$_3$Si$_2$Te$_6$ collapsing into one single curve where $T_C$ is located at the zero point of the horizontal axis.

### E. $1 \times 18$ H$^+$/cm$^2$ (H ∥ ab) Mn$_3$Si$_2$Te$_6$ Critical Analysis and Scaling Relations

Figure S5 (a), (b), and (c) show the converged plots for the KF method with critical exponents $\beta$ and $\gamma$, the $M_S$ and $\chi_0^{-1}$ power law fits with respective $T_C$'s, and the MAP with a linear fit showing the best fit for this critical regime, respectively for the proton irradiation at $1 \times 10^{18}$ H$^+$/cm$^2$. Figure S3 (d) and (e) show the scaling relations $M|\varepsilon|^{-\beta}$ vs. $H|\varepsilon|^{-(\gamma+\beta)}$ and $m^2$ vs. $h/m$, respectively where two branches of data should form, one below and one above the $T_C$. The insets of Figure S5 (d) and (e) show the log $(M|\varepsilon|^{-\beta})$ vs. log $(H|\varepsilon|^{-(\gamma+\beta)})$ where two branches should fall into one, and the $MH^{-1/\delta}$ vs. $\varepsilon H^{-1/(\beta\delta)}$ where the zero point of the horizontal axis indicates the $T_C$, respectively. From these scaling relation plots we confirm that the magnetic critical behavior analysis we performed for the $1 \times 10^{18}$ H$^+$/cm$^2$ MST sample is true and properly scaled, thus yielding reliable critical exponents $\beta$, $\gamma$ and $\delta$.

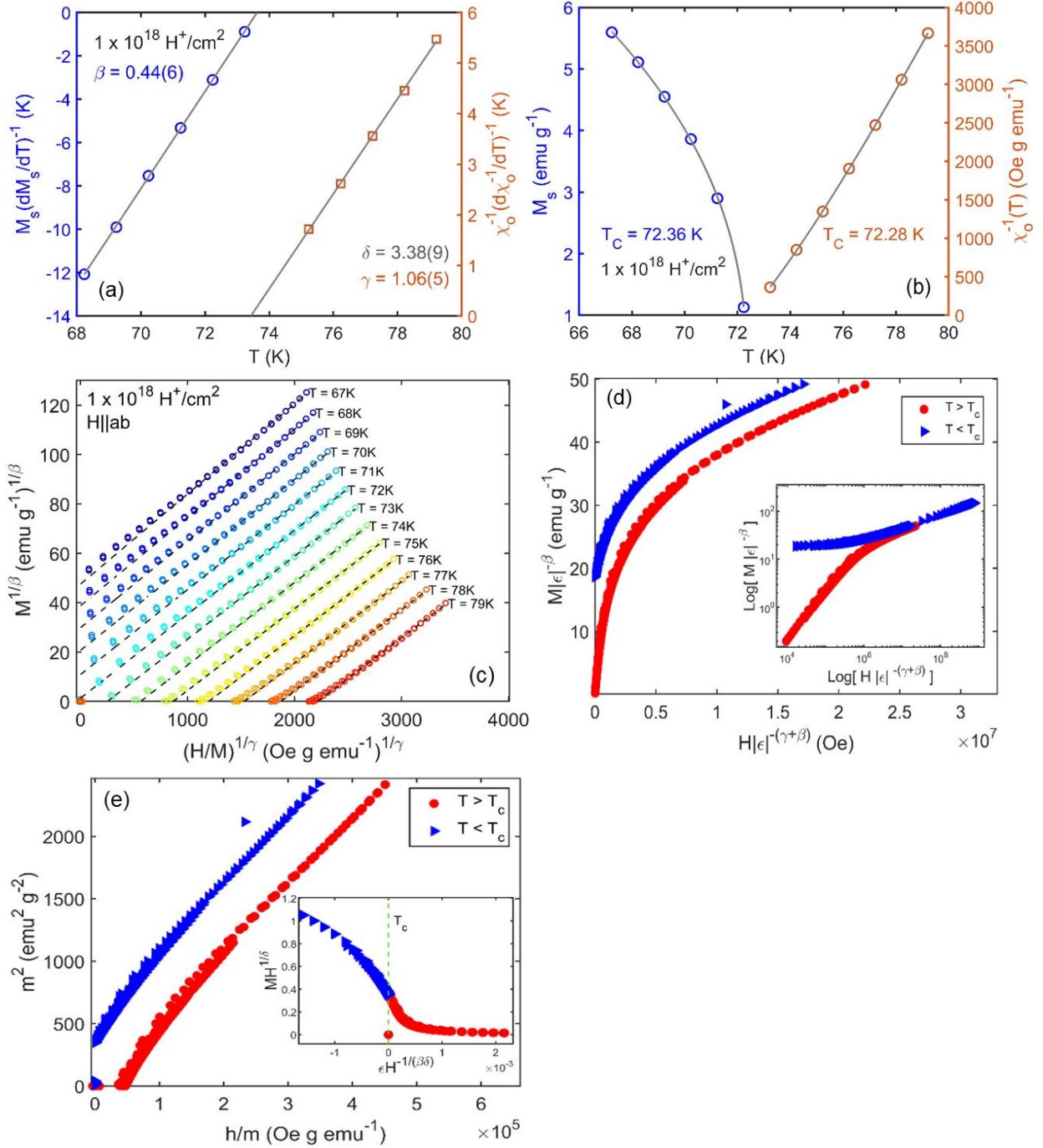

FIG. S5. (a) Kouvel-Fisher plot for $1 \times 10^{18}$ H$^+$/cm$^2$ Mn$_3$Si$_2$Te$_6$ with critical exponents $\beta$, $\gamma$, and $\delta$ using the Widom scaling relation. (b) Magnetization and susceptibility analysis. (c) MAP for $1 \times 10^{18}$ H$^+$/cm$^2$ Mn$_3$Si$_2$Te$_6$ with a linear fit on isotherms with KF critical exponents. (d) Scaling of renormalized magnetization, $m$, and renormalized field, $h$, with isotherms falling into two branches, below and above $T_C$ $1 \times 10^{18}$ H$^+$/cm$^2$ Mn$_3$Si$_2$Te$_6$. Inset shows the log-log scale of the renormalized $m$ and $h$ for the $1 \times 10^{18}$ H$^+$/cm$^2$ case. (e) Data from $1 \times 10^{18}$ H$^+$/cm$^2$ H $\parallel ab$ Mn$_3$Si$_2$Te$_6$ isotherms collapse into two separate branches, one below and one above $T_C$, which further confirm the reliability of the exponents in the critical regime.

Inset shows experimental data for $1 \times 10^{18}$ H$^+$/cm$^2$ Mn$_3$Si$_2$Te$_6$ collapsing into one single curve where $T_C$ is located at the zero point of the horizontal axis.

## F.  Critical exponents and $T_C$ as a function of proton irradiation

Figure S6 (a) shows $\beta$ and $\gamma$ plotted as a function of proton irradiation. It is observed that $\beta$ is largest for 1 x 10$^{15}$ H$^+$/cm$^2$ and $\gamma$ largest for 5 x10$^{15}$ H$^+$/cm$^2$. Similarly in Figure S6 (b) $\delta$ is largest for the fluence rate of 5 x10$^{15}$ H$^+$/cm$^2$. Figure S6 (c) displays the $T_C$ obtained from KF analysis plotted as a function proton irradiation where see minimal decreases in the transition temperature for the fluences 5 x 10$^{15}$ and 1 x 10$^{18}$ H$^+$/cm$^2$.

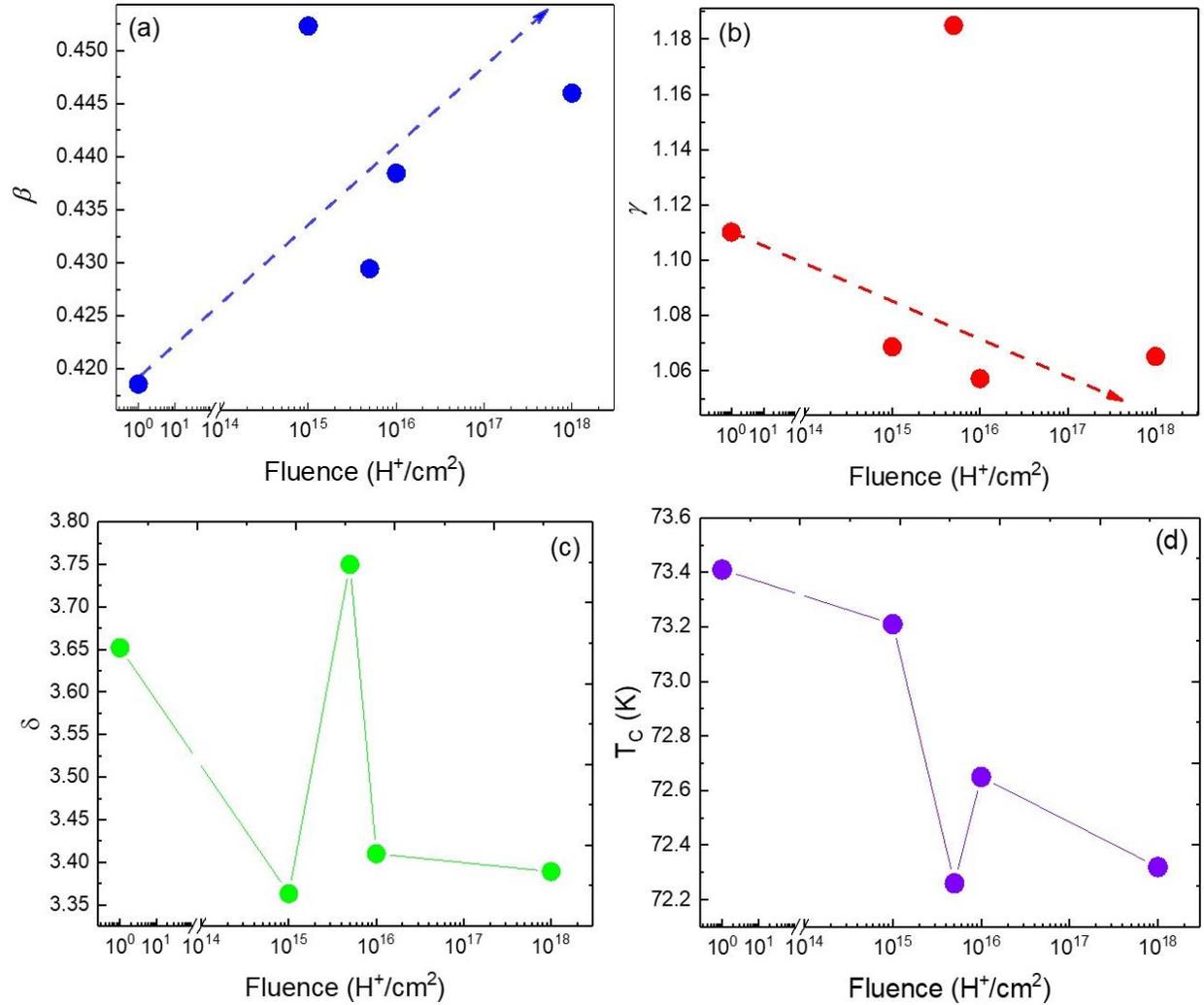

FIG. S6. (a) $\beta$, $\gamma$, and (b) $\delta$ exponent values and (c) Curie temperature from Kouvel-Fisher method plotted as a function of fluence.

## G.  Negative change in magnetic entropy

Figure S7 presents the negative change in magnetic entropy $(-\Delta S_M)$ for pristine and proton irradiated samples along the easy axis of MST at magnetic fields of 1, 1.5, 2, 2.5, and 3 T. The maximum values correspond to the area in which magnetic transition occurs. At the maximum applied magnetic field (3 T), we clearly observe that the maximum $-\Delta S_M$ is the largest for the fluence $5 \times 10^{15}$ H$^+$/cm$^2$.

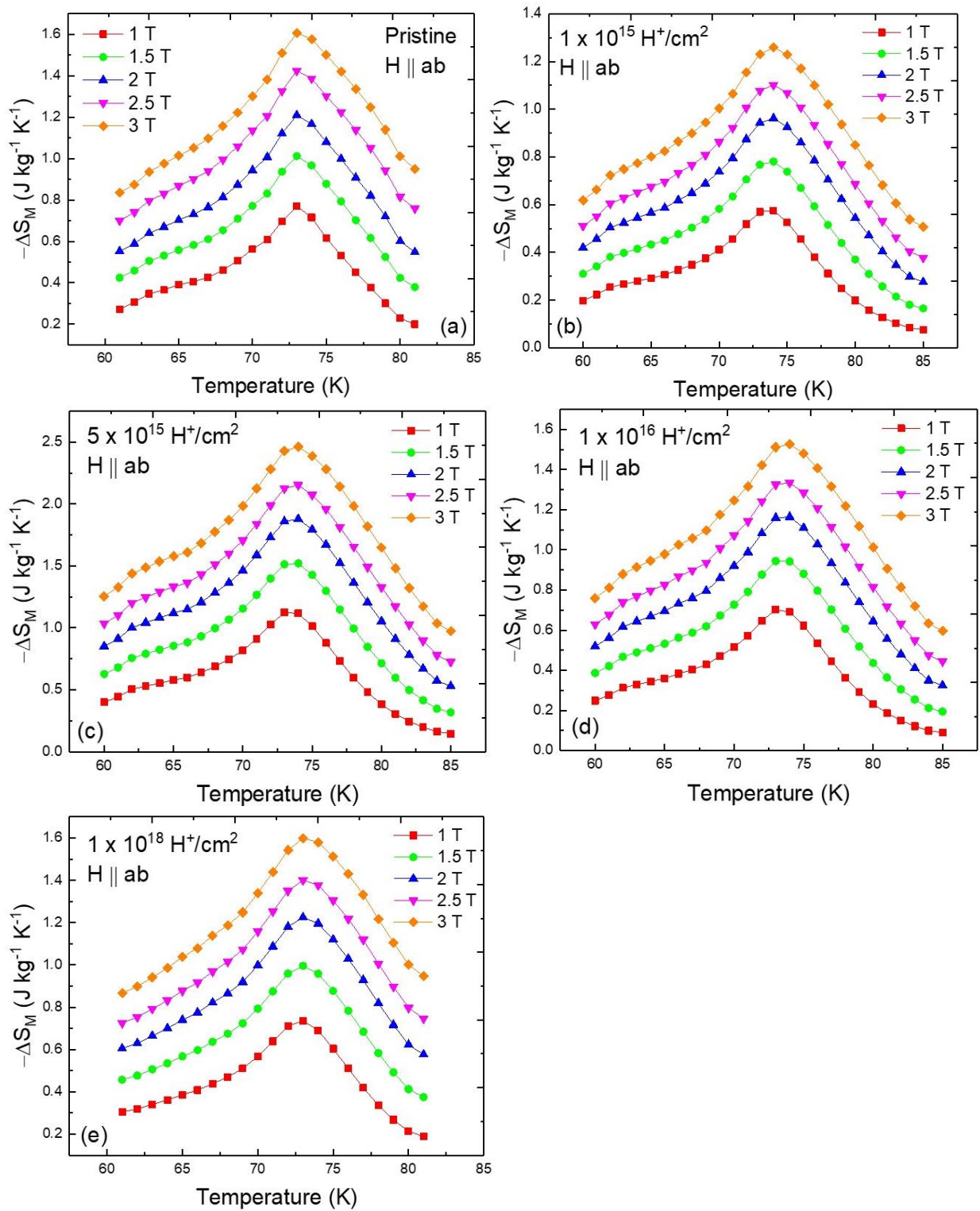

FIG. S7. The negative magnetic change in entropy for select fields up to 3 T for each fluence (a) pristine, (b) 1 x 10¹⁵, (c) 5 x 10¹⁵, (d) 1 x 10¹⁶, and (e) 1 x 10¹⁸ H⁺/cm².

### H.  Magnetic entropy through heat capacity analysis

Heat capacity measurements are performed for all pristine and proton irradiated samples. Two forms of analysis are performed utilizing heat capacity to determine the magnetic entropy; employing a linear background subtraction and through applying an Einstein fit to obtain the magnetic contribution(s). Table S1 presents the magnetic entropy ($S_M$) for the linear background subtraction method. The largest $S_M$ for this method is for 1 x 10¹⁵ H⁺/cm². The percentage(s) in Table S1 represents an estimate of how much entropy was released in the magnetic transition in comparison to total spin (or magnetic) entropy that should be released as spins go from disordered in the paramagnetic phase to ordered in the long-range magnetic ordering phase. Here we see that these percentages are much smaller than they realistically should be, supporting the idea that we have frustrated system for the MST pristine and proton irradiated samples. Figure S8 displays the heat capacity data with the linear background that is to be subtracted to obtain the magnetic entropy for (a) 1 x 10¹⁶ and (b) 1 x 10¹⁸ H⁺/cm². Similar is performed for the other fluences. Table S2 presents the $S_M$ for the method utilizing Einstein fit(s) on the heat capacity data for pristine and proton irradiated samples of MST. Figure S9 shows the heat capacity data treated with Einstein fittings and the magnetic contribution for (a) pristine and (b) 1 x 10¹⁵ H⁺/cm².

Table S1. Magnetic entropy for pristine and irradiated MST for linear background subtraction method.

| Fluence | $S_M$ ( J/K) | Percentage |
|---|---|---|
| Pristine | 0.00199 | 13 % |
| 1 x 10¹⁵ H⁺/cm² | 0.0024 | 16 % |
| 5 x 10¹⁵ H⁺/cm² | 0.0016 | 11 % |
| 1 x 10¹⁶ H⁺/cm² | 0.00187 | 12 % |
| 1 x 10¹⁸ H⁺/cm² | 0.00187 | 12 % |

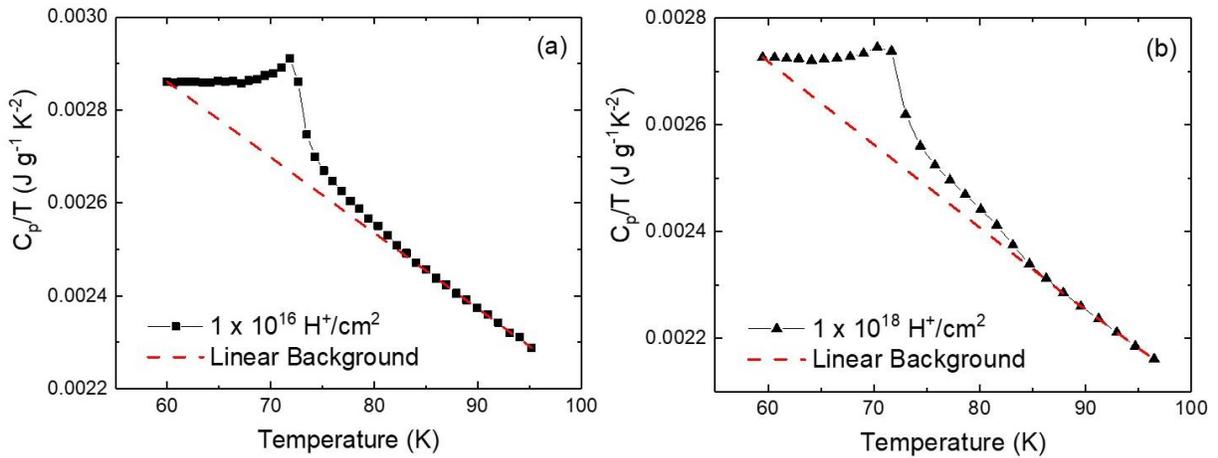

FIG. S8. Linear background subtraction plots of the heat capacity shown for the two fluences (a) 1 x 10$^{16}$ and (b) 1 x 10$^{18}$ H$^+$/cm$^2$.The linear background that was subtracted to obtain to magnetic entropy is seen as the red dashed line.

Table S2. Magnetic entropy for pristine and proton irradiated samples of MST utilizing Einstein fits.

| Fluence | $S_M$ ( J/K) |
|---|---|
| Pristine | 0.0041 |
| 1 x 10$^{15}$ H$^+$/cm$^2$ | 0.0037 |
| 5 x 10$^{15}$ H$^+$/cm$^2$ | 0.0035 |
| 1 x 10$^{16}$ H$^+$/cm$^2$ | 0.0028 |
| 1 x 10$^{18}$ H$^+$/cm$^2$ | 0.0031 |

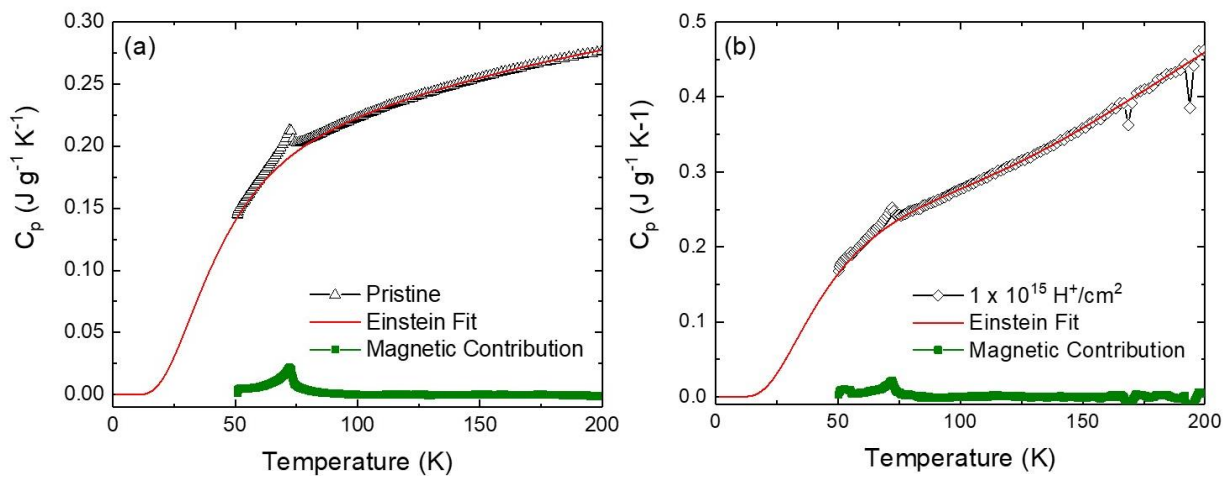

FIG. S9. Heat capacity data fitted with an Einstein fit (red curve) for the (a) pristine sample and (b) 1 x 10$^{15}$ H$^+$/cm$^2$. Magnetic contribution of the data is seen in green.

## SUPPLEMENTAL MATERIAL REFERENCES